\documentclass{article}

\usepackage{subcaption}
\usepackage{amsmath}
\usepackage{xcolor}
\usepackage{PRIMEarxiv}
\usepackage{nameref}
\usepackage[utf8]{inputenc} 
\usepackage[T1]{fontenc}    
\usepackage{hyperref}       
\usepackage{url}            
\usepackage{booktabs}       
\usepackage{amsfonts}       
\usepackage{nicefrac}       
\usepackage{microtype}      
\usepackage{lipsum}
\usepackage{fancyhdr}       
\usepackage{graphicx}       
\graphicspath{{media/}}     

\title{Opinion Dynamics on Complex Networks
}

\author{
  Jiarui Dong\\
  School of Science,\\
  Beijing University of Posts and Telecommunications,\\
  Beijing, China
  \AND
  Yi-Cheng Zhang \\
  Department of Physics,\\
  University of Fribourg,\\
  Fribourg, Switzerland
  \And
  Yixiu Kong  \\
  School of Science,\\
  Beijing University of Posts and Telecommunications,\\
  Beijing, China\\
  \texttt{yixiu.kong@bupt.edu.cn}
}

\begin{document}
\maketitle

\begin{abstract}
Social media has emerged as a significant source of information for people. As agents interact with each other through social media platforms, they create numerous complex social networks. Within these networks, information spread among agents and their opinions may be altered by their neighbors' influence. This paper explores opinion dynamics on social networks, which are influenced by complex network structure, confirmation bias, and specific issues discussed. We propose a novel model based on previous models to simulate how public opinion evolves from consensus to radicalization and to polarization. We also analyze how agents change their stance under their neighbors' impact. Our model reveals the emergence of opinion groups and shows how different factors affect opinion dynamics. This paper contributes to the understanding of opinion dynamics on social networks and their possible applications in finance, marketing, politics, and social media. 
\end{abstract}


\section{Introduction}
The application of complexity theory to various fields has increased with the advancement of complex networks in recent years\cite{erdos1960evolution,watts1998collective,barabasi1999emergence,boccaletti2006complex}. The methodology of complexity has been applied in many fields. One of the fields that benefit from this approach is social sciences, where complex networks can help reveal the underlying mechanisms of different social phenomena\cite{castellano2009statistical,siegel2009social,jusup2022social}. 
Social media have emerged as a significant and integral aspect of social life that have transformed the modes of information acquisition and communication\cite{carr2015social}. Social media users are not only passive recipients of information but also active producers\cite{obar2015social}. Therefore, they are involved in a dynamic process of influencing and being influenced by other users. Based on these features, a complex network model can be constructed to represent the behavior of social media users. In this model, each node corresponds to a user and each link represents a relationship between users.

The study of opinion dynamics aims to understand how collective opinion patterns change and individual opinion develops\cite{lorenz2007continuous, acemoglu2011opinion, cao2012overview, dong2018survey, noorazar2020recent}. In this paper, we put forward a novel model based on the model proposed by Fabian et al.\cite{baumann2020modeling}.  By combining complex network theory and opinion dynamics model, this paper presents a novel model based on the one proposed by Fabian et al.\cite{baumann2020modeling}. The original model is modified in a reasonable and realistic way, which incorporates the network structure and other necessary factors. Some intriguing phenomena that were overlooked in \cite{baumann2020modeling} are also identified and explained with reasonable hypotheses.

The structure of this paper is as follows. Section \ref{sec:Related Works} reviews the development of opinion dynamics models. Section \ref{sec:Results} describes the model and the experimental design and presents the results from perspectives of both collective opinion and individual opinion. Section \ref{sec:Conclusion} concludes the paper and discusses the limitations and future directions of the proposed model.

\section{Related Works}
\label{sec:Related Works}
This section provides an overview of some influential models in the field of opinion dynamics. Some notations are defined for the clarity of the subsequent discussion. The social network composed of $N$ agents is denoted by $G$. We define an adjacency matrix as $A$ and $A_{ij}=1$ when there are relationships between agent $i$ and $j$.
Referring to different problems and different models, agent's opinion can be defined in continuous opinion space such as $\mathbb{X}=\left [ 0, 1 \right ]$ or discrete opinion space like $\mathbb{X} =\left \{ 0, 1 \right \}$. The opinion of an agent at time t can be expressed as $x_i^{(t)}$.

With time elapsing, the agents' opinion evolves and collective opinions also come to different states either when we stop the process manually or the system converges spontaneously. In our context, we observe several kinds of states of collective opinions. If the majority of agents hold similar neutral or moderate opinions, we name this state as a \textit{consensus}. In contrast, if most of them take a similar radical stance we call this state \textit{radicalization}. In addition to the states mentioned above, we use the term \textit{polarization} when there are only two main groups of agents with opposing opinions. While \textit{Fragmentation} means more than two opinion groups exist.

We begin by introducing some models that are defined in a discrete opinion space. Next, we discuss some models that are defined in a continuous opinion space, which motivated us to develop our own model.
\paragraph{Discrete-space-opinion Models}
One of the most classic discrete-space-opinion models is the voter model introduced by Holley and Liggett\cite{holley1975ergodic}. As it describes, every agent placed in a regular lattice takes a binary stance which is denoted as a binary variable. At a given time $t$, a random picked agent will adjust his opinion according to his one randomly selected neighbors' opinion. The study of voter models has been extended to different network topology\cite{suchecki2004conservation, suchecki2005voter, sood2005voter, sood2008voter}. In addition, there are some extensions in the aspect of agents' opinion update rules. Vazques and Redner\cite{vazquez2004ultimate} proposed a constrained 3-state voter model, where agents can be leftists, centralists or rightists. Wang et el. combined voter model with bounded confidence model and define the model in a continuous opinion space\cite{wang2020public}. Horstmeyer and Kuehn brought up a model which is another breakthrough of adaptive voter model like\cite{benczik2009opinion}. They deployed adaptive voter model on simplicial complexes\cite{horstmeyer2020adaptive}.

Discrete-space-opinion models are not only dominated by the voter model. Sznajd model conceptualizes an one-dimensional lattice consisting $N$ agents. Their opinions can be either $\left\{ -1, 1\right\}$. At time $t$, if randomly picked agent $i$ and $i+1$s' opinions satisfy $x_{i}^{(t)}x_{i+1}^{(t)}=1$ then agent $i-1$ and $i+2$ take the opinions of agent $i$ and $i+1$. Otherwise, agent $i-1$ and $i+2$ takes agent $i$ and $i+1$'s  opinions respectively\cite{sznajd2000opinion, sznajd2005sznajd}.
\paragraph{Continuous-space-opinion Models}
In this section we start with the DeGroot model\cite{degroot1974reaching}, which assumes a constant transition matrix $W$. The evolution of agents' opinions can be written as\begin{equation}
     X^{(t)}=WX^{(t-1)}
 \end{equation}
 where $X^{(t)}$ is the vector of agents' opinions at time $t$ and if not specified, $X^{(t)}\subseteq \mathbb{R}^N$. $W$ can be regarded as a weight matrix, which manifests how an agent's neighbors are influential to him.
 As the DeGroot model is linear, in \cite{berger1981necessary} the authors provide a necessary and sufficient condition for the DeGroot model' s reaching consensus. Considering agents' preconception, Friedkin and Johnsen extents the DeGroot model like
 \begin{equation}
     X^{(t)}=AWX^{(t-1)}+(I-A)X^{(1)}
 \end{equation}, in which $A=diag(a_{11},a_{22},\cdots ,a_{NN})$ denotes agents' susceptibilities to other's impact and $0<a_{ii}<1$.

Next we introduce bounded confidence models (BC models) , which offer more flexibility  than the DeGrootian models. In the BC models, agents react indifferently to agents whose opinions are too divergent from theirs. In the DW model\cite{deffuant2000mixing}, if two randomly selected agents have $\left|x^{(t)}_i-x^{(t)}_j \right|\leq \varepsilon $ their opinions will update as 
\begin{equation}
    \begin{cases}
        x_i^{t+1}=x_i^{t}+\mu (x_j^{t}-x_i^{t})\\
        x_j^{t+1}=x_j^{t}+\mu (x_i^{t}-x_j^{t})
    \end{cases}
\end{equation}
where $\mu\in [0,0.5] $ measures to what degree agents $i$ and $j$ are impact each other. Another classic BC model is the HK model\cite{hegselmann2002opinion}. We update agent $i$'s opinion as
\begin{equation}
    x^{(t+1)}_j=\left| I(i,x^{(t)}))\right|^{-1}\sum_{j\in I(i,x^{(t)}))}x_j^{(t)}
\end{equation}
where $I(i,x)=\begin{Bmatrix}
1< j\leq n\mid \left|x_i-x_j \right|\leq \varepsilon 
\end{Bmatrix}$, which is similar to the DW model.The HK model and the DW model differ in how agents adjust their opinions based on social influence. In the HK model, agents consider the average opinion of a larger group of agents, whereas in the DW model, agents are influenced by a random pairwise interaction with a neighbor.

Previously we have introduced homogeneous BC models above, which means $\varepsilon $ and $\mu $ are uniform for every agent. Otherwise, we call the models heterogeneous\cite{hegselmann2002opinion, kou2012multi}. Bounded confidence models have other variations when other effects are considered. Jalili studied the effects of social leaders and social pressure in these models\cite{jalili2013effects}. 
Li et al. proposed a HK model that considered peer pressure\cite{li2018opinion}. Kurahashi-Nakamur et al. attempted to explain patterns of opinion diversity when agents were influenced by others who held distant views\cite{kurahashi2016robust}. However, these interactions rarely occurred.

Last, we introduce the model proposed by Baumann et el.\cite{baumann2020modeling, baumann2021emergence}. For a system with $N$ agents, each agent holds an time-varying opinion $x^{(t)}_i\in \left ( -\infty ,+\infty  \right )$. $\sigma(x_i) $ describes the
agent’s qualitative stance towards a binary issue and $\left|x_i \right|$ quantifies the agent's conviction of his stance. Agents' opinions evolve according to activity-driven (AD) interactions. Each agent $i$ has an activity level $a_i\in [\varepsilon ,1]$, which represents the propensity they will interact with $m$ random agents. $a$ follows a power-law distribution
\begin{equation}\label{eq:(5)}
    F(a)=\frac{1-\gamma }{1-\varepsilon^{1-\gamma }}a^{-\gamma }
\end{equation} where $\gamma$ controls the power law decay and $\varepsilon$ is the minimum activity. 
If agent $i$ is activated he will choose whether interact with agent $j$ based on homophily. In other words, follow possibility
\begin{equation}\label{eq:(6)}
p_{ij}=\frac{\left|x_i-x_j \right|^{-\beta }}{\sum_{j}\left|x_i-x_j \right|^{-\beta }}
\end{equation}, in which $\beta$ includes various homophily effects.
Then we can describe AD temporal networks with adjacency matrix $A(t)$, where $A_{ij}(t)=1$ means agent $i$ influences agent $j$ at a given time $t$. Considering reciprocity, agent $j$ can also influence agent $i$ following possibility $r$.
And now we have agent's opinion dynamics
\begin{equation}\label{eq:(7)}
    \dot{x}=-x_i+K\sum_{j=1}^{N}A_{ij}(t)tanh(\alpha x_j)
\end{equation}
where $K>0$ is the social interaction strength and $\alpha$ is characterized as a given issue's controversialness which means controlling the nonlinear function's shape.
\section{Results}
\label{sec:Results}
In this section, we first propose our opinion dynamics model based on Baumann model. In \cite{baumann2020modeling} only the collective opinion pattern was studied. However, various details can be observed despite the collective opinion. To further explore the phenomena and mechanism, we discuss opinion dynamics from both individual and collective perspectives.
\subsection{Models and Methods}
In Baumann’s model, an activated agent can influence $m$ distinct agents. This implies that any agent can influence any other agent through social media regardless of their social distance. However, this assumption is unrealistic. For example, on Twitter, agents may be influenced by tweets from people they do not follow (as Baumann’s model suggests). But according to Ref. \cite{nickerson1998confirmation}, agents tend to follow people who share similar opinions with them.  Ref. \cite{dunbar1992neocortex} also suggests that although agents encounter many like-minded people through recommendation systems, they only form a few stable connections. Therefore, in our context, we assume that agents have stronger relations with their followers and followees. Agents are more likely to be influenced by them and to influence them in return. To model this situation, we first consider that agents have stable social relations represented by a static complex network.  Then we use a time-varying matrix $A(t)$ to denote how they influence each other at a given time $t$. Thus, our model generalizes Baumann’s model, which assumes a fully-connected network to represent agents’ relatively stable social relations.

Mathematically, the opinion dynamics is described as follows:
\begin{equation}
    \dot{x}_i=-x_i+\sum_{j=1}^{N}c_jA_{ij}(t)tan\left ( \alpha x_j \right )
\end{equation}.
First we assume a static network $S$ with $N$ agents, which represents agents' stable social relations. Here, $c_i$ is the closeness centrality\cite{bavelas1950communication} of agent $i$ in the social network. Closeness centrality is a commonly used way to measure the efficiency of a node's spreading information in a network\cite{zhang2017degree}. We replace the original parameter $K$ in Baumann’s model with each agent’s personal centrality. We do this because we think that the concept of social interaction strength is vague and allows for artificial manipulation. In contrast, centrality is an intrinsic property of agents, which may be determined by their personality, social status or other traits that affect their position in the social network. This way, we can also account for the network structure, which is ignored in the original model.

As Ref. \cite{baumann2020modeling} states, Baumann's model is based on a minimal number of assumptions.  Therefore, we also do not extend the former model by considering features that are observed in empirical social networks but not essential and fundamental (such as targeted advertising and user credibility mentioned by Baumman et el.) and that may affect how agents’ opinions evolve. To preserve the property of least assumptions, we incorporate agents’ social relations (network structures) by simply replacing the social strength $K$ with the node centrality $c_i$ in the social network.

Centrality is an important indicator corresponding to node positions in complex networks. There exist many kinds of definitions of centrality which depict node's characteristics from various perspectives. In our paper, we use \textit{closeness} centrality. Freeman first explored this centrality both conceptually and experimentally\cite{freeman1979centrality, freeman2002centrality}, which can be calculated as follows
\begin{equation}
    c_i=\frac{N-1}{\sum_{\forall j,i\neq j}d(j,i)}
\end{equation}
where $d(j,i)$ is the length of shortest path from $i$ to $j$ and $N$ is the number of nodes in a network. It is worth noting that \textit{closeness} centrality above can only be applied to networks without a disconnected component. But this limitation can be easily overcome by using the extended version of disconnected networks\cite{murray1965improved}. In many practical applications, information flows along the shortest path\cite{saxena2020centrality}. In our paper, agents directly receive and generate influence 
from and on their nearest neighbors. However, this is not denying that agents are immune from the influence of people who are not directly linked to them. In the light of our opinion dynamics model, information is not restricted to flow along a single type of trajectory (say geodesics). Borgatti analyzed the typology of flow processes and the most appropriate centrality measures fitting different scenarios\cite{borgatti2005centrality} conceptually. Borgatti listed some typical traffics like E-mail broadcast, attitude influencing, and so on, which are better measured using \textit{closeness} centrality. Therefore, \textit{closeness} centrality is used in our paper, while other centralities remain possible candidates for introducing the network structure into the model. 

We preserve the activity-driven process. As we mentioned, $A(t)$ is not an adjacency matrix of a temporal network. At a given time $t$, an activated (refer to Eq. \ref{eq:(5)}) agent can influence his neighbors with a probability given by Eq. \ref{eq:(6)}. Thus, we represent $A(t)$ as a transient influence relationship matrix among agents compared to the static social network we defined formerly.

\subsection{Collective opinion patterns}\label{sec:3.2}


\begin{figure}
     \centering
     \begin{subfigure}[b]{0.48\textwidth}
         \centering
         \includegraphics[width=\textwidth]{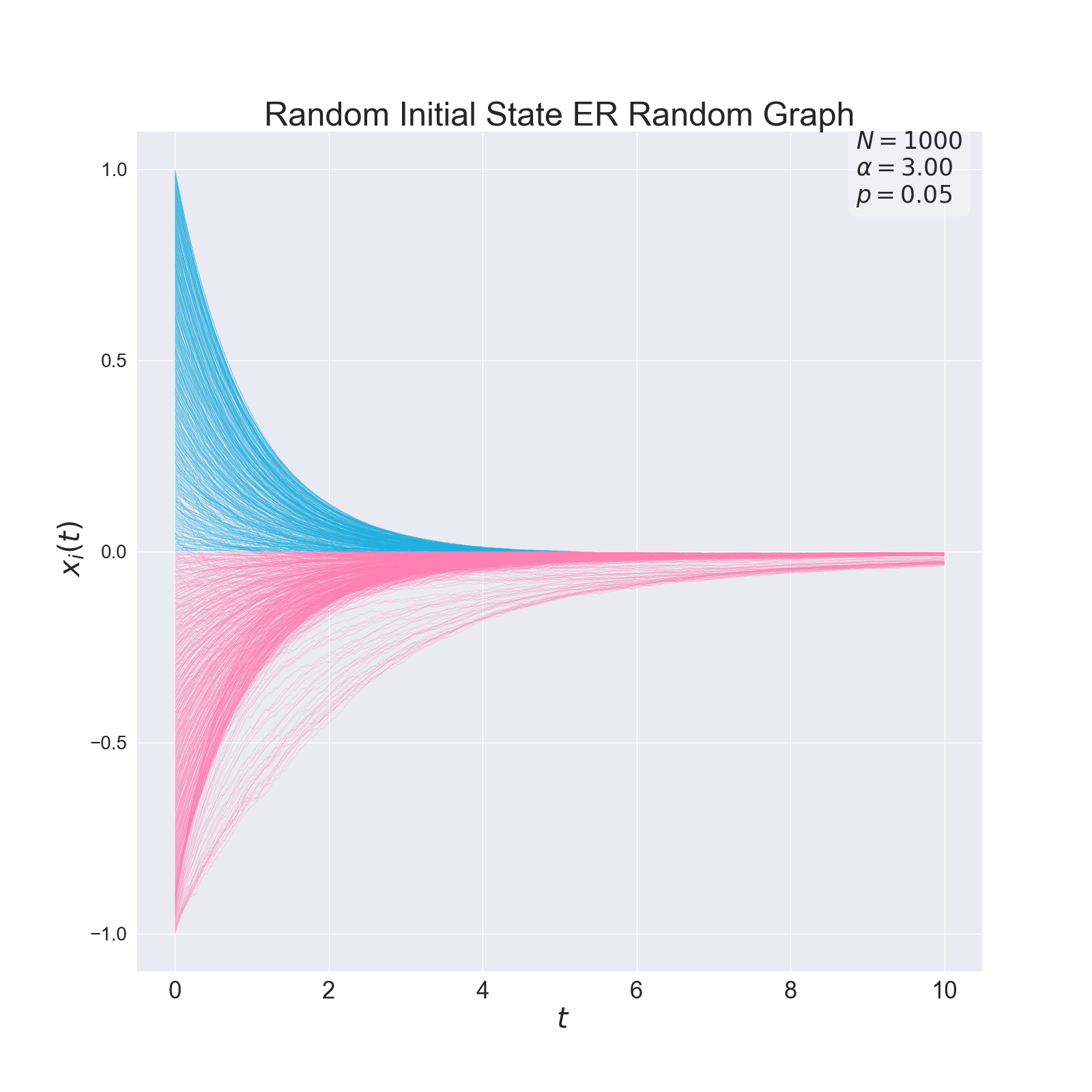}
         \caption{\textit{Neutral Consensus}}
         \label{fig:n1}
     \end{subfigure}
     \hfill
     \begin{subfigure}[b]{0.48\textwidth}
         \centering
         \includegraphics[width=\textwidth]{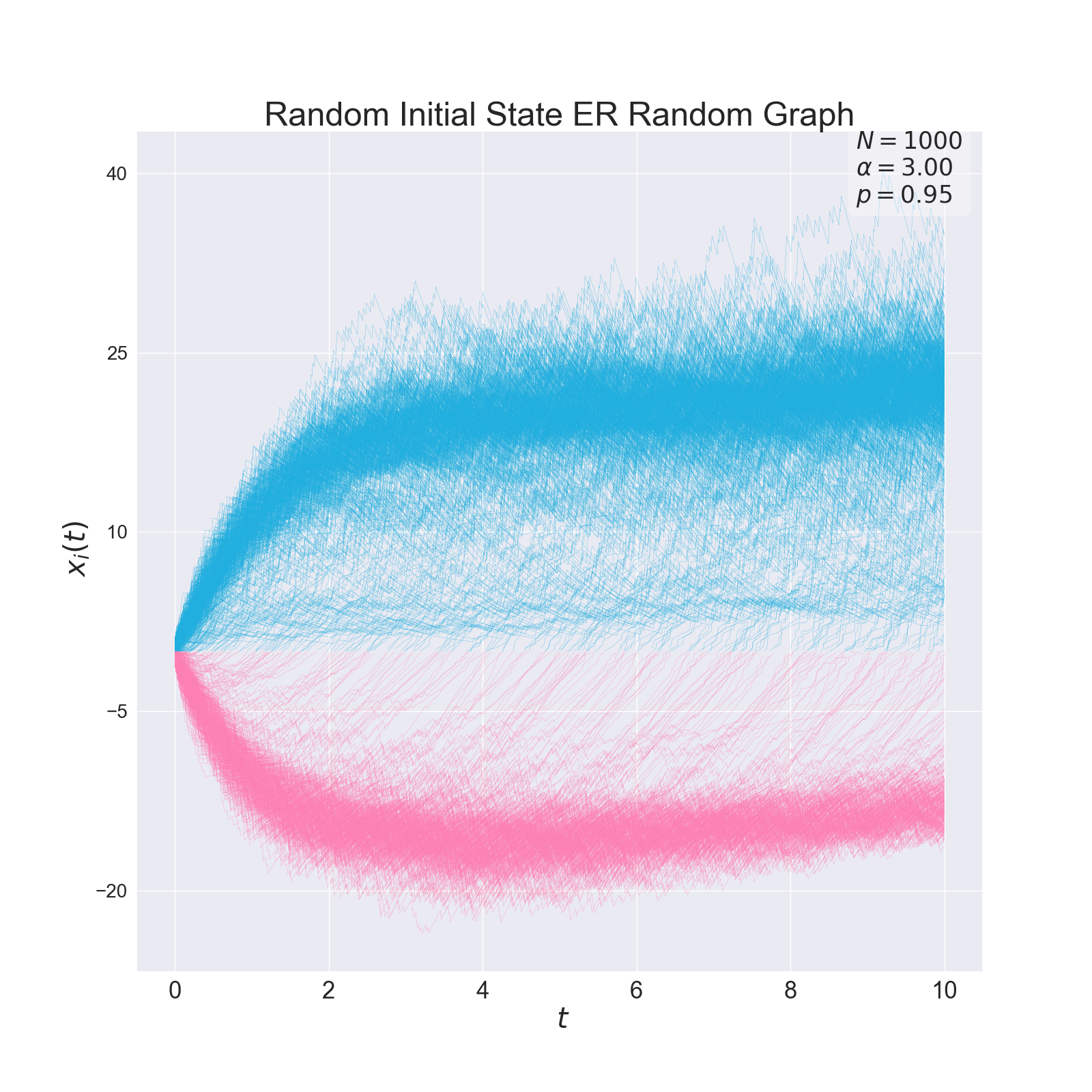}
         \caption{\textit{Polarization}}
         \label{fig:n2}
     \end{subfigure}
     \hfill
     \begin{subfigure}[b]{0.48\textwidth}
         \centering
         \includegraphics[width=\textwidth]{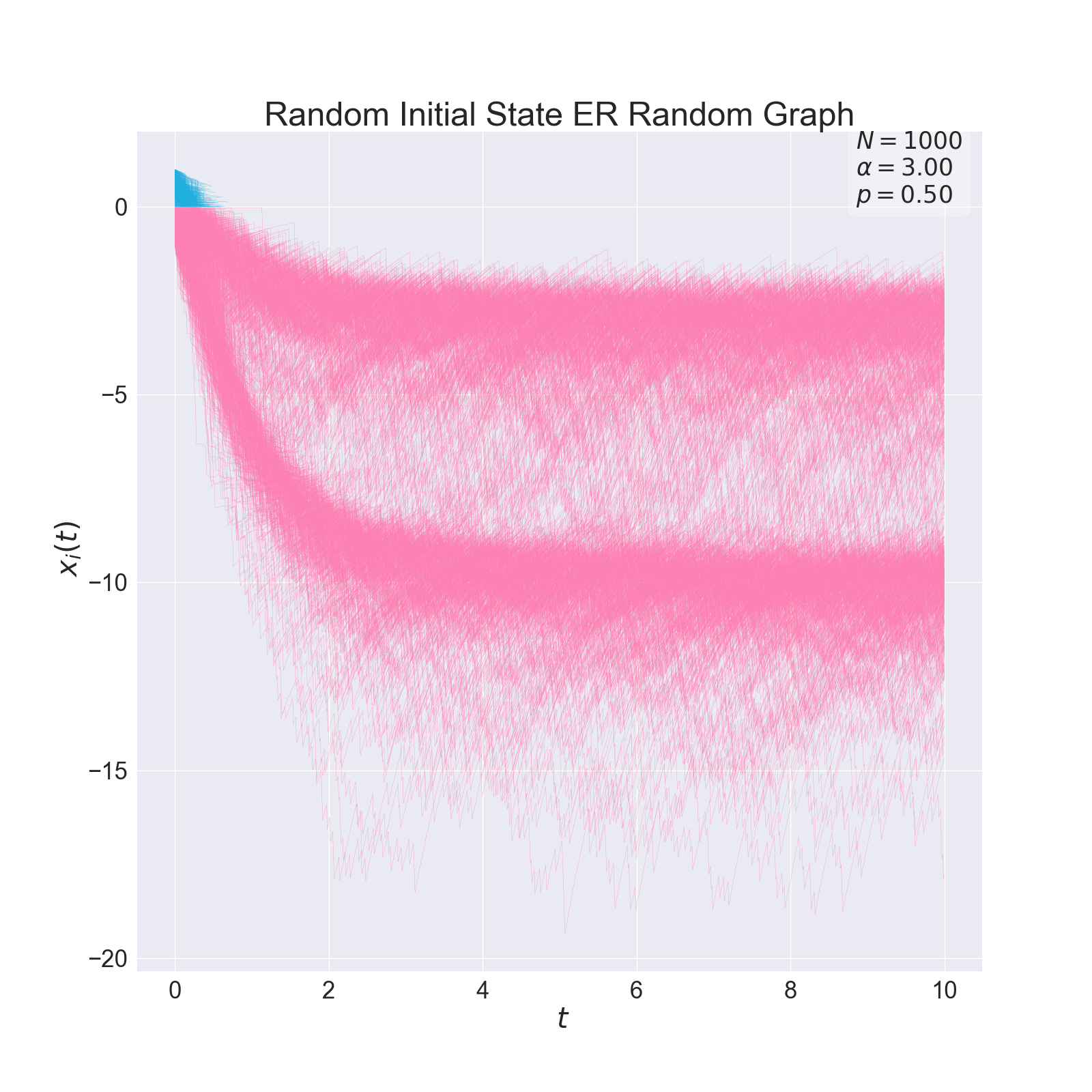}
         \caption{\textit{Radicalization}}
         \label{fig:n3}
     \end{subfigure}
     \hfill
        \caption{Three major collective opinion patterns. Positive opinions are marked as blue while negative opinions are marked as pink. We set $\beta =1$, $\alpha =3$, $dt =0.01$ for every simulation. Fig. \ref{fig:n1}comes to the \textit{consensus} when most agents' opinions converge to zero. In Fig. \ref{fig:n3}, we can see most agents take the same stance. This state is \textit{radicalization}. In Fig. \ref{fig:n2}, agents split into two dichotomous groups and we obtain an opinion \textit{polarization}.}
        \label{fig:1}
\end{figure}


\begin{figure}
     \centering
     \begin{subfigure}[b]{0.48\textwidth}
         \centering
         \includegraphics[width=\textwidth]{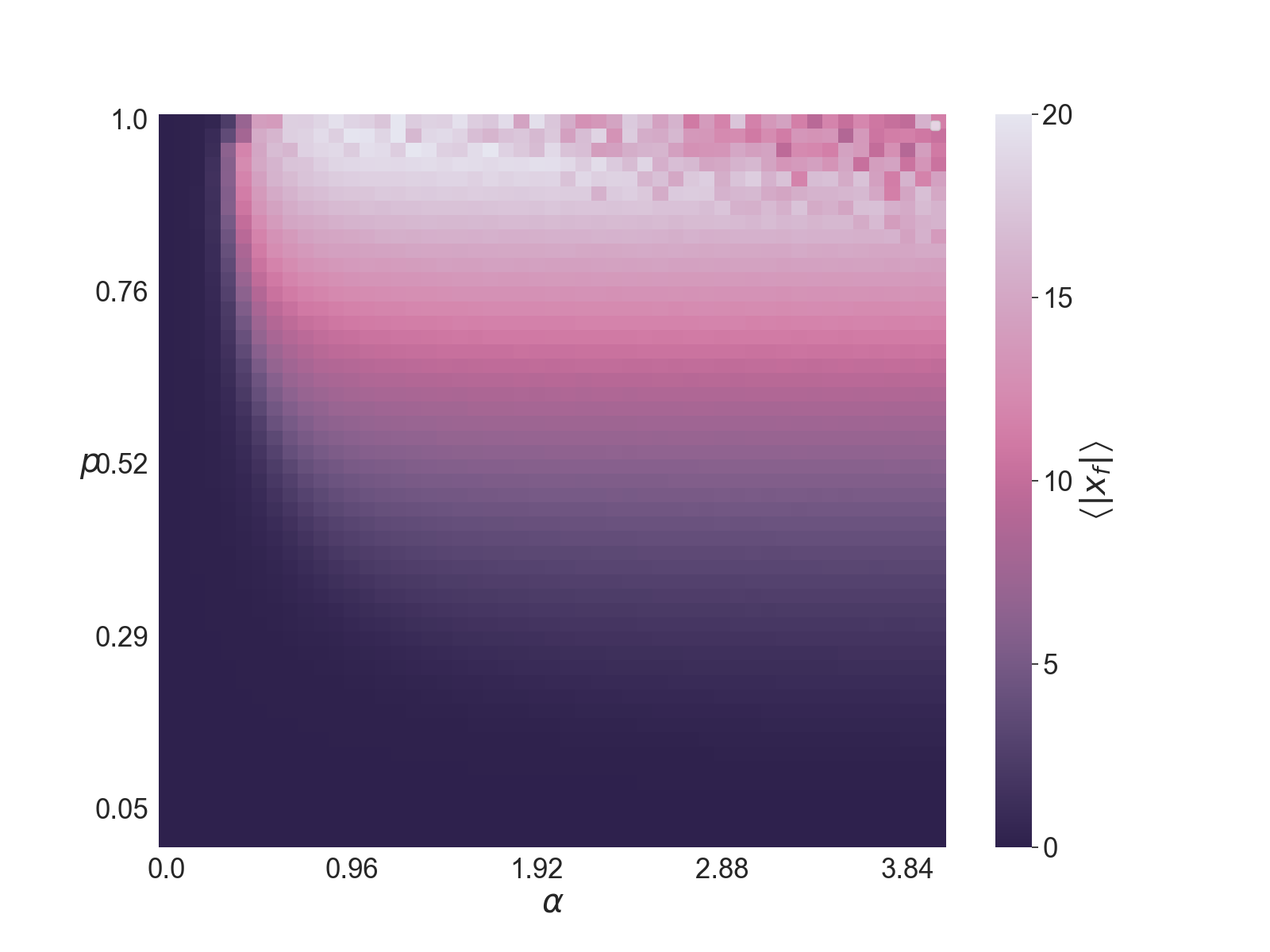}
         \caption{$\left<\left| x_f\right| \right>$}
         \label{fig:n4}
     \end{subfigure}
     \hfill
     \begin{subfigure}[b]{0.48\textwidth}
         \centering
         \includegraphics[width=\textwidth]{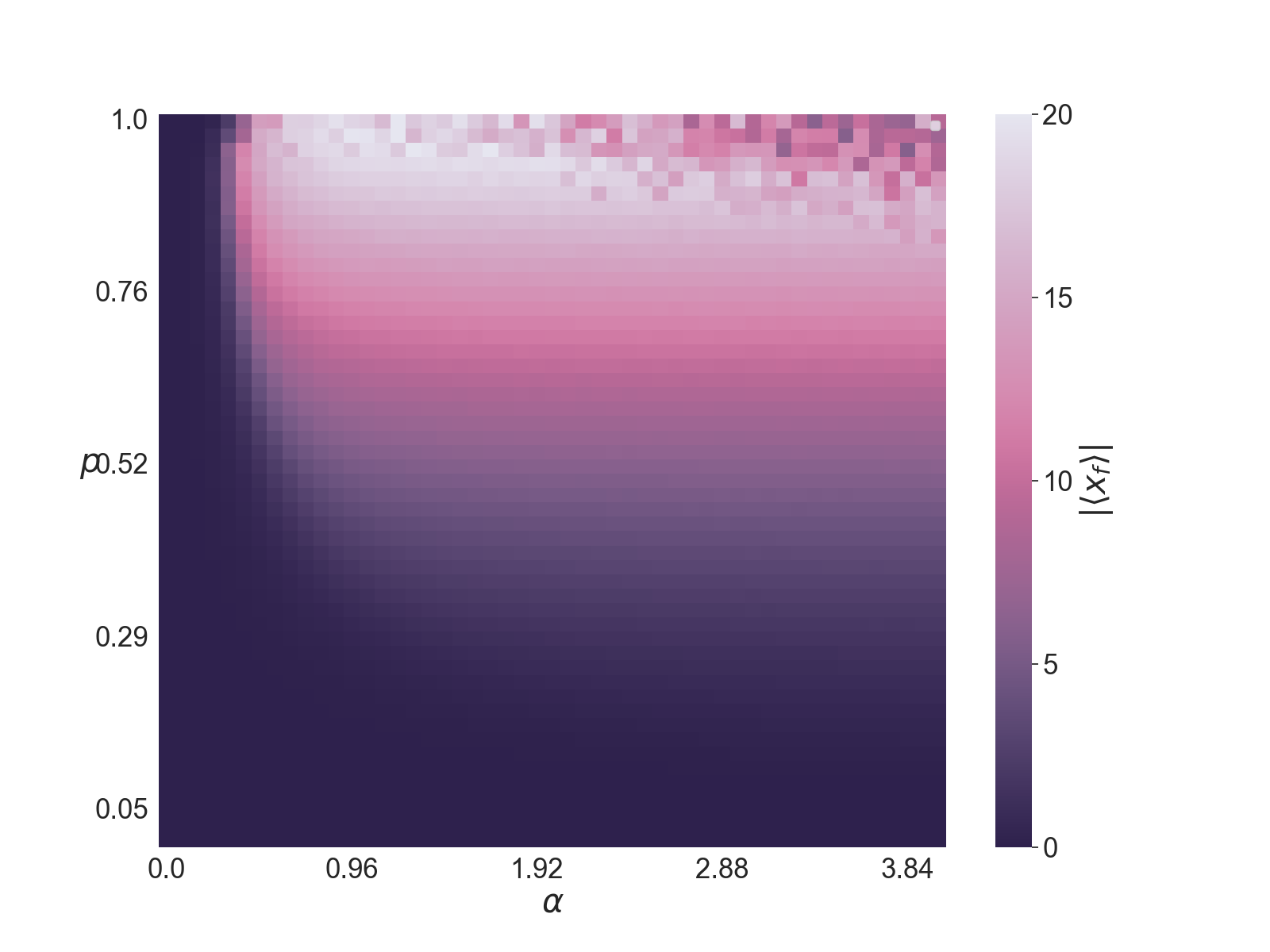}
         \caption{$\left|\left<x_f \right> \right|$}
         \label{fig:n5}
     \end{subfigure}
     \hfill
     \begin{subfigure}[b]{0.48\textwidth}
         \centering
         \includegraphics[width=\textwidth]{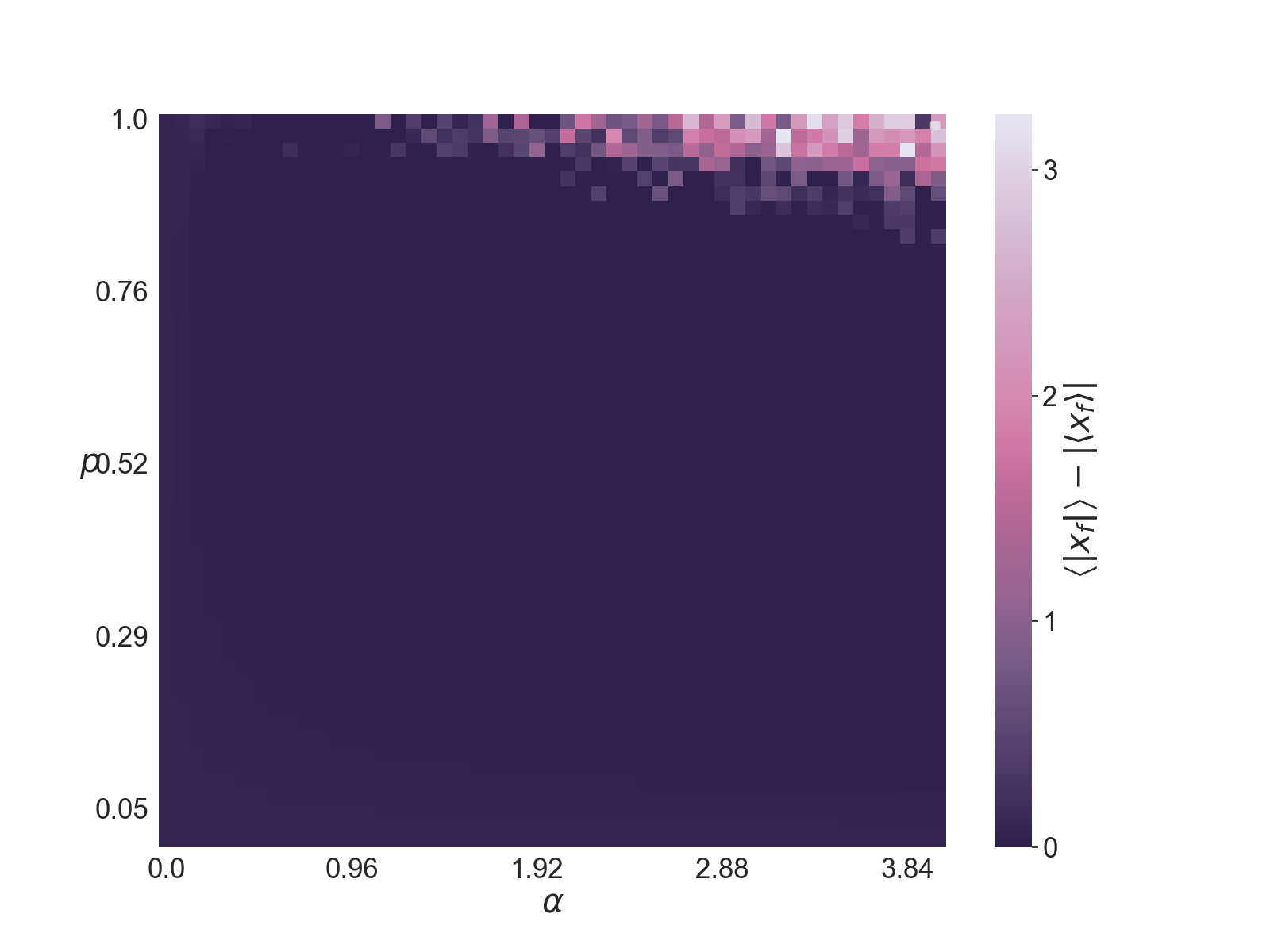}
         \caption{$\left|\left<x_f \right> \right|-\left<\left|x_f \right| \right>$}
         \label{fig:n6}
     \end{subfigure}
     \hfill
     \begin{subfigure}[b]{0.48\textwidth}
         \centering
         \includegraphics[width=\textwidth]{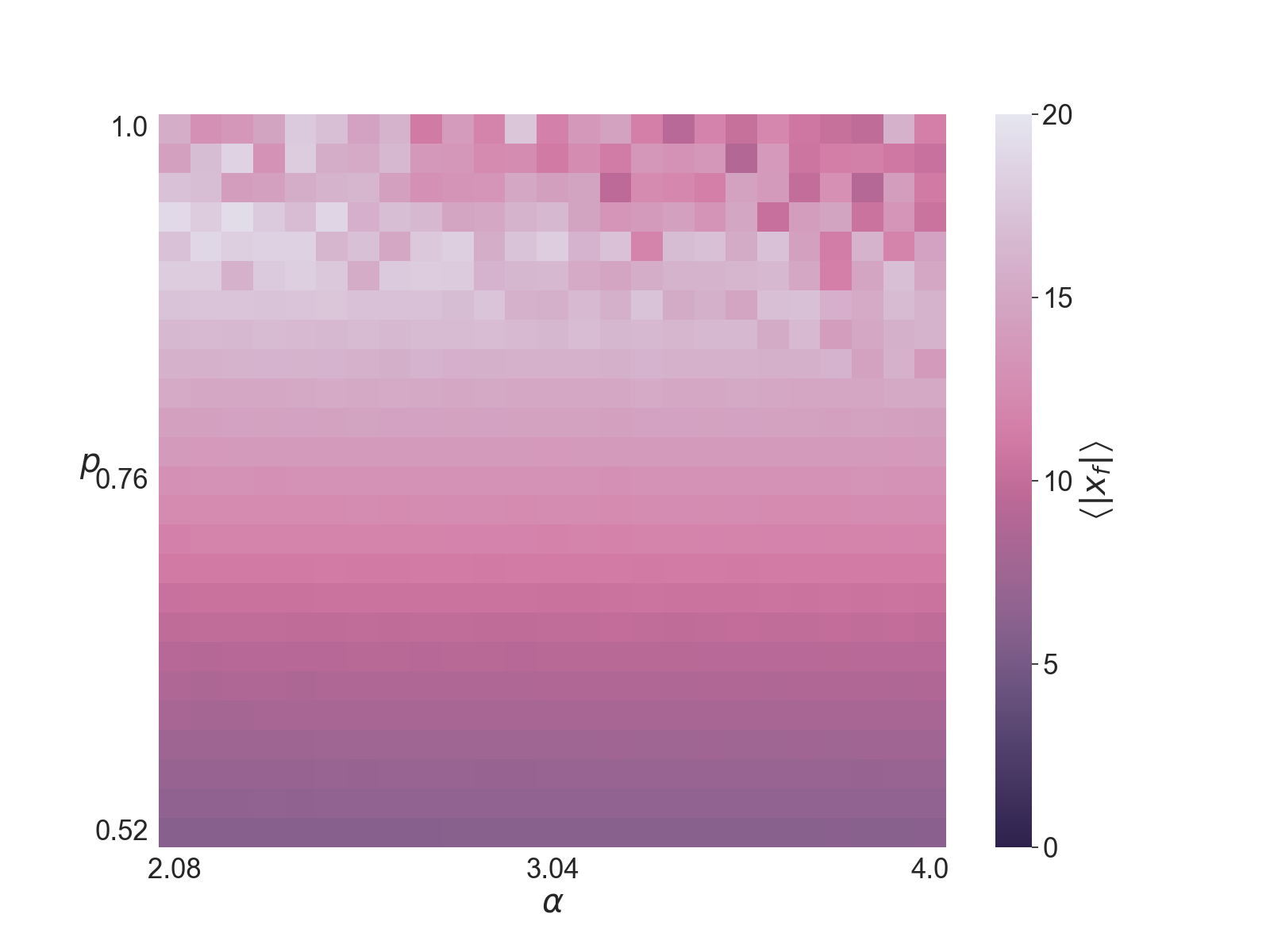}
         \caption{Local zoom of the miscellaneous part in Fig. \ref{fig:n4}}
         \label{fig:n6__}
     \end{subfigure}
        \caption{Transition from consensus to radicalization and to polarization. Parameter settings are in accordance with Fig. \ref{fig:1} shows except of $\rho$ and $\alpha $. 
    The results of every parameter tuple are averaged after 10 runs. We show absolute values of the average final opinions $\left|\left<x_f \right> \right|$ and average values of the absolute final opinions $\left<\left| x_f\right| \right>$ in Fig. \ref{fig:n4} and Fig. \ref{fig:n5} respectively. In the dark regions, agents reach a consensus. In the light regions in the middle, a radicalization happens and agents become more radical with the increase of $\alpha $ and $\rho $. While agents are polarized in the miscellaneous region on the top. Fig. \ref{fig:n6} shows $\left|\left<x_f \right> \right|-\left<\left|x_f \right| \right>$ in the phase space leaving the polarization states only.}
        \label{fig:2}
\end{figure}

The patterns of collective opinion and its transition are evident when people with similar views form groups and isolate themselves from opposing opinions. This phenomenon, known as echo chambers, has been empirically analyzed in different social networking platforms using various methods \cite{del2016echo, cota2019quantifying, cinelli2020covid, cinelli2021echo}. The authors also verified their model by studying some widely discussed issues on Twitter\cite{baumann2020modeling}.

We start our numerical simulations in the following scheme. We generate a static random network $S$ with $N$ nodes, representing agents' stable social relations. We control the network structure by changing the average degree $\left<k \right>$ (We achieve this by tuning the probability $\rho $ of establishing a random link. We can initialize the agent's opinion with a random value in the interval $[-1, 1]$, thus, $x_i^{(0)}\in [-1,1]$. In this context, we initialize the opinion variable $x$ uniformly in space $[-1, 1]$ like \cite{baumann2020modeling}. Unlike the model in\cite{baumann2020modeling}, we do not limit the number of agents that an activated agent can influence. On the contrary, an activated agent can influence any number of agents from his neighbors in $S$.

In Fig.\ref{fig:1} we show three different patterns of collective opinion and detailed parameter settings are listed. We will not discuss the effect of homophily at this part, so we set $\beta$ as a constant now. As the effect of controversialness has been studied in \cite{baumann2020modeling}, we also set $\alpha$ constant. Compared with the results of Baumann's, we observe when the issue is controversial but static social network $S$ on which agents reside, is sparse, agents also come to a \textit{consensus}. Though agents may become radical and have a stronger desire to communicate with others meeting a controversial issue they will gradually become wavering spontaneously for the lack of reinforcement of comrades. With $\rho$ increasing to 0.5 (see Fig. \ref{fig:n3}) , an \textit{radicalization} happens. Differently, we can spot an obvious stratification in the dominant group, which is called \textit{fragmentation}. This phenomenon was neglected in the original model. That is to say, there are some diehards but also some kind of moderate people in a radical group. In the next section, we will discuss this phenomenon.  If we make $S$ extremely dense (say $\rho=1$), we have the same assumption as the Baumann model.


Consequently, we find in Fig. \ref{fig:n2} a \textit{polarization} is reached. But we still perceive some interesting phenomena here. During the temporal evolution of agents' opinions, we can see some negative agents change their stance. This is more realistic because the crowds are not stagnant. People's thoughts are definitely always changing. The only difference is whether they become firmer or wavering. We will further discuss why these renegades behave like this in the next section. In addition, we notice the original model has robustness respecting to the asymmetry of initial conditions and the effect of fluctuations. In other words, the collective pattern is predetermined by the parameters they set. Surprisingly, even when negative agents are very few compared to positive agents, they still reach the same outcome as when both groups are equal. This seems to ignore people’s mobility. In our paper, different initial conditions lead to different collective opinion patterns.

\begin{figure}
    \centering
    \includegraphics[scale=0.25]{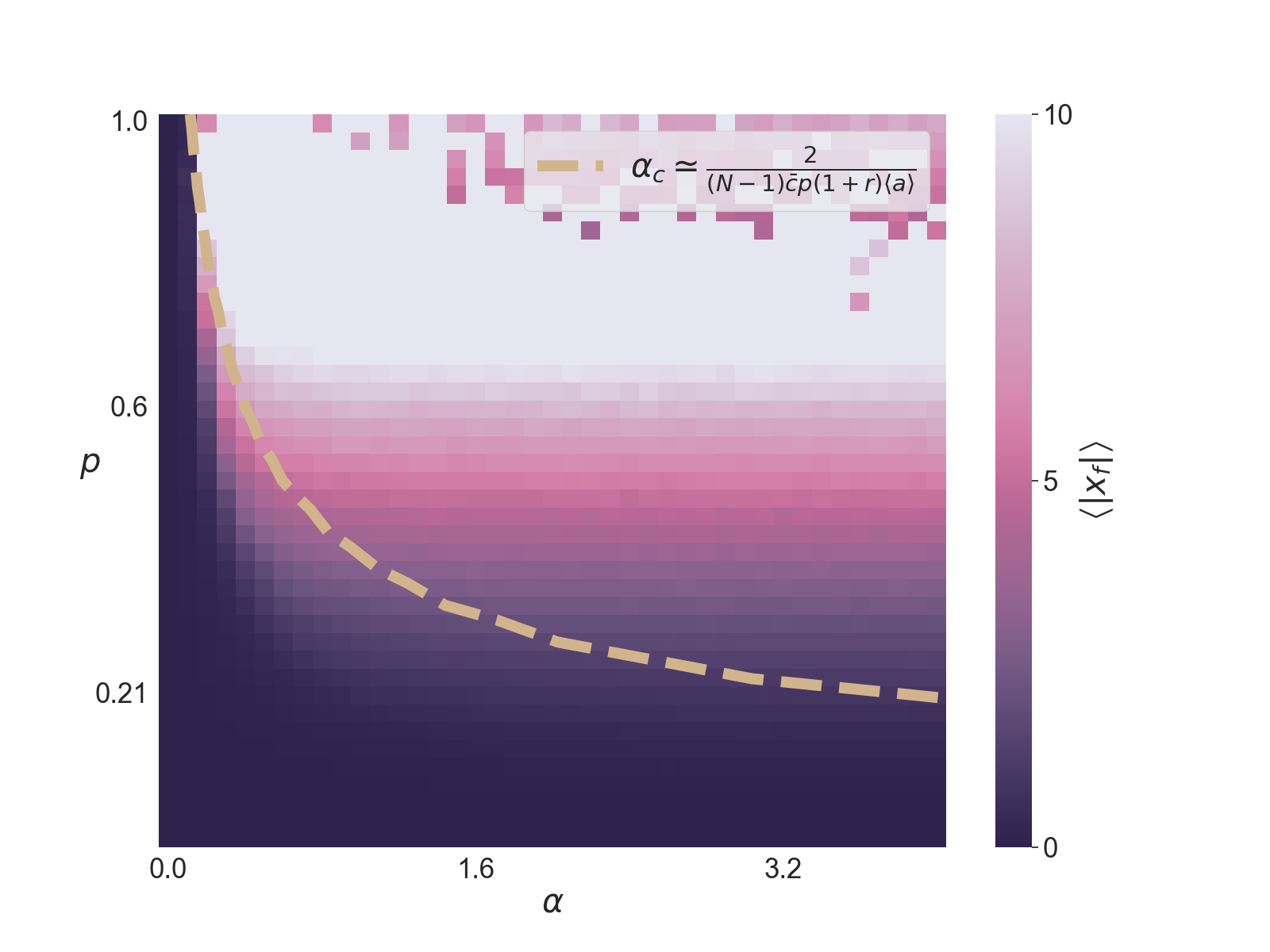}
    \caption{Criticality between consensus and radicalization}
    \label{fig:3}
\end{figure}

Next, we show the phase transition of collective opinions from \textit{consensus} to \textit{radicalization} and to \textit{polarization} in Fig.\ref{fig:2}. If we only calculate the absolute values of the average final opinions $\left|\left<x_f \right> \right|$ , it may lead to an ambiguity of the state of \textit{polarization} and \textit{consensus} because the agents' opinions counteract in \textit{polarization} states. So, we both calculate separately the absolute values of the average final opinions $\left|\left<x_f \right> \right|$ and average values of the absolute final opinions $\left<\left| x_f\right| \right>$. In the dark areas, when $\rho $ is small, $\alpha $ can make little contribution to agents' radicalism, we can only observe \textit{consensus}. As $\rho $ and $\alpha $ increase, agents' opinions become more radical. And in the top miscellaneous region, we notice bright areas and dark areas coexist. To figure out what happens under these settings of $\rho $ and $\alpha $, we plot $\left|\left<x_f \right> \right|-\left<\left|x_f \right| \right>$ on the phase space. This is a simple idea. At the state of \textit{consensus}, agents' are neutral. In other words, $\left|x_f \right|=0$, thus, we have $\left|\left<x_f \right> \right|-\left<\left|x_f \right| \right>=0$. Similarly, when it comes to \textit{radicalization}, most agents take the same stance. That's to say, $\sigma(x_f) $ are the same. We can also have $\left|\left<x_f \right> \right|-\left<\left|x_f \right| \right>=0$. But at the state of \textit{polarization}, conflicting opinions will counteract when calculating $\left|x_f \right|$, thus $\left|\left<x_f \right> \right| \neq \left<\left|x_f \right| \right>$. In consequence, at the miscellaneous part, agents approach a \textit{polarization} (also see Fig.\ref{fig:n6}) . For better illustration, we also zoom the miscellaneous part of Fig. \ref{fig:n4} in Fig. \ref{fig:n6__}. The transition from \textit{radicalization} to \textit{polarization} has not been discussed in the Ref. \cite{baumann2020modeling}. Besides, we calculate an analytical expression of critical $\alpha $ with the mean-field assumption (setting homophily $\beta = 0$). We give the analytical expression of critical points as
\begin{equation}
    \alpha_c\simeq \frac{2}{(N-1)\bar{c}p(1+r)\left<a \right>} 
\end{equation}
where $\bar{c}$ is the average closeness centrality of agents in $S$ and other parameters have been mentioned above. See pink star dashed line in Fig. \ref{fig:3}, which is in perfect accordance with numerical simulations.

\subsection{Individual opinion evolution}
\begin{figure}
    \centering
    \includegraphics[scale=0.5]{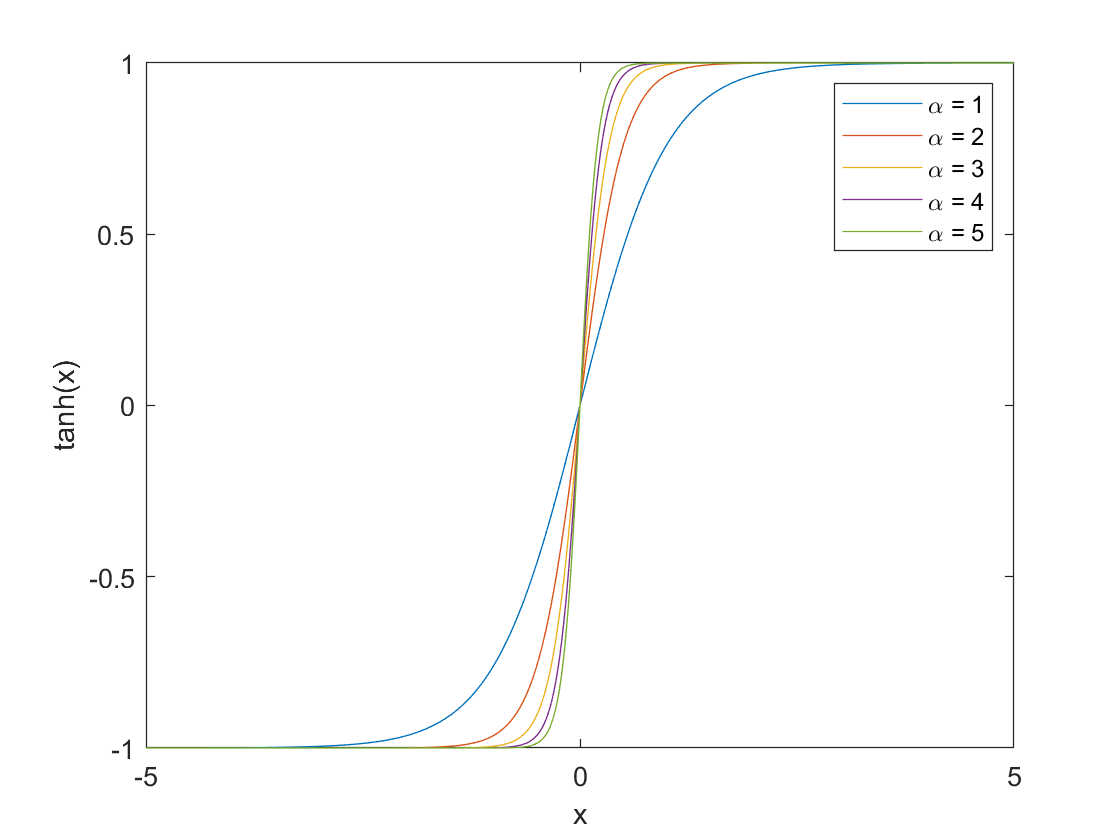}
    \caption{$tanh(x)$ with different $\alpha$}
    \label{fig:4}
\end{figure}

In this section, we study how individual opinion evolves. We take a look at the nonlinear function $tanh(x)$ at first. Check Fig. \ref{fig:4}, we find when $\alpha $ is larger than 2 (just qualitatively), further increase $\alpha $ seems to have little influence on the shape of function.  To account for the issue controversialness, when $\left|x^{(t)} \right|$ is large, we need an odd and sensitive nonlinear function instead of $tanh(x)$.  One such function could be:
\begin{equation}
    f(x)=   \begin{cases}
                &ln(\alpha x+1) \text{ if } x\geq 0 \\
                &-ln(-\alpha x+1) \text{ if } x< 0 
            \end{cases}
\end{equation}.
Using this alternative function, we will see how key opinion groups emerge and agents give up on their original stance during temporal opinion evolution.

\begin{figure}[h]
    \centering
    \includegraphics[scale=0.25]{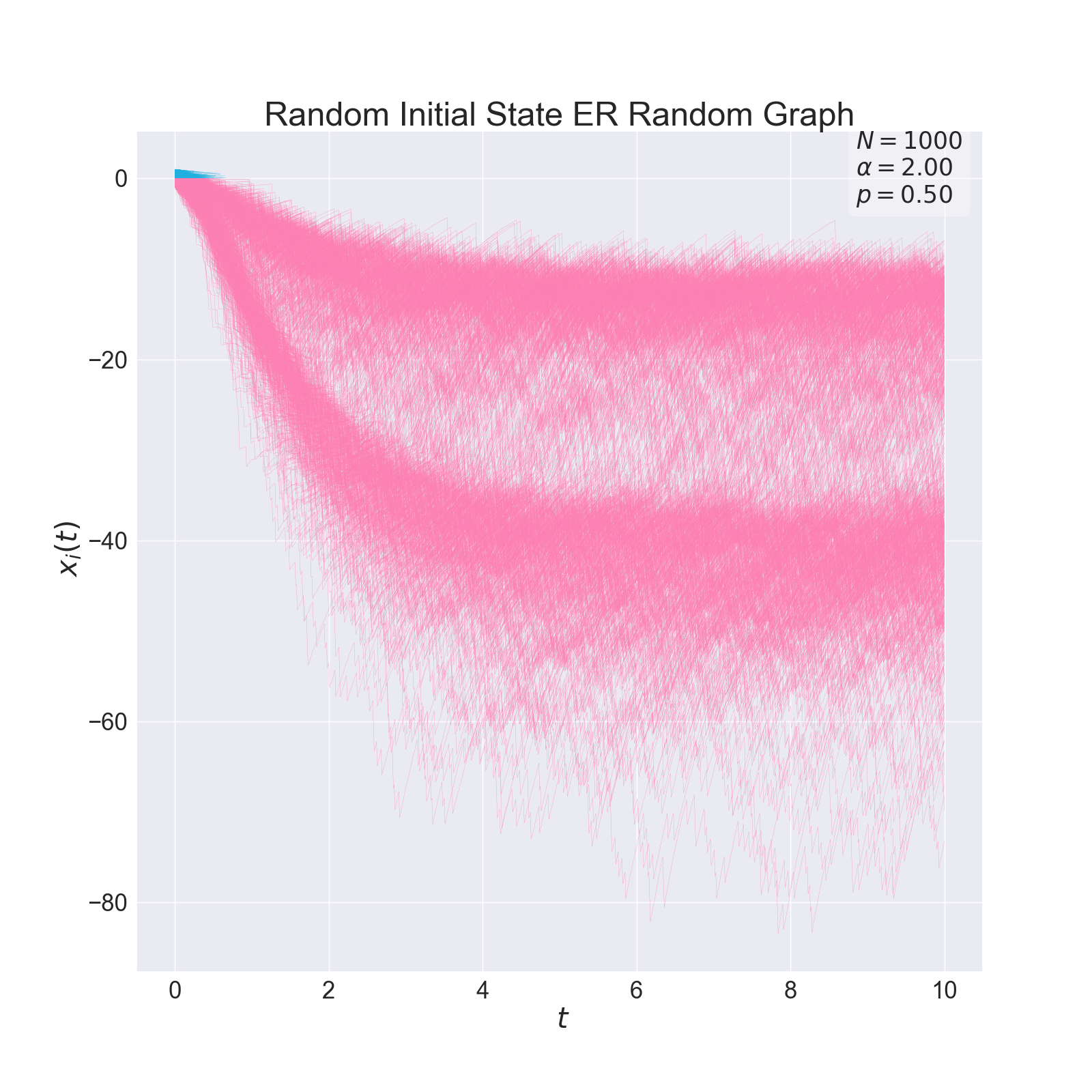}
    \caption{Fragmentation in temporal evolution of agents' opinions}
    \label{fig:5}
\end{figure}

\begin{figure}
     \centering
     \begin{subfigure}[b]{0.48\textwidth}
         \centering
         \includegraphics[width=\textwidth]{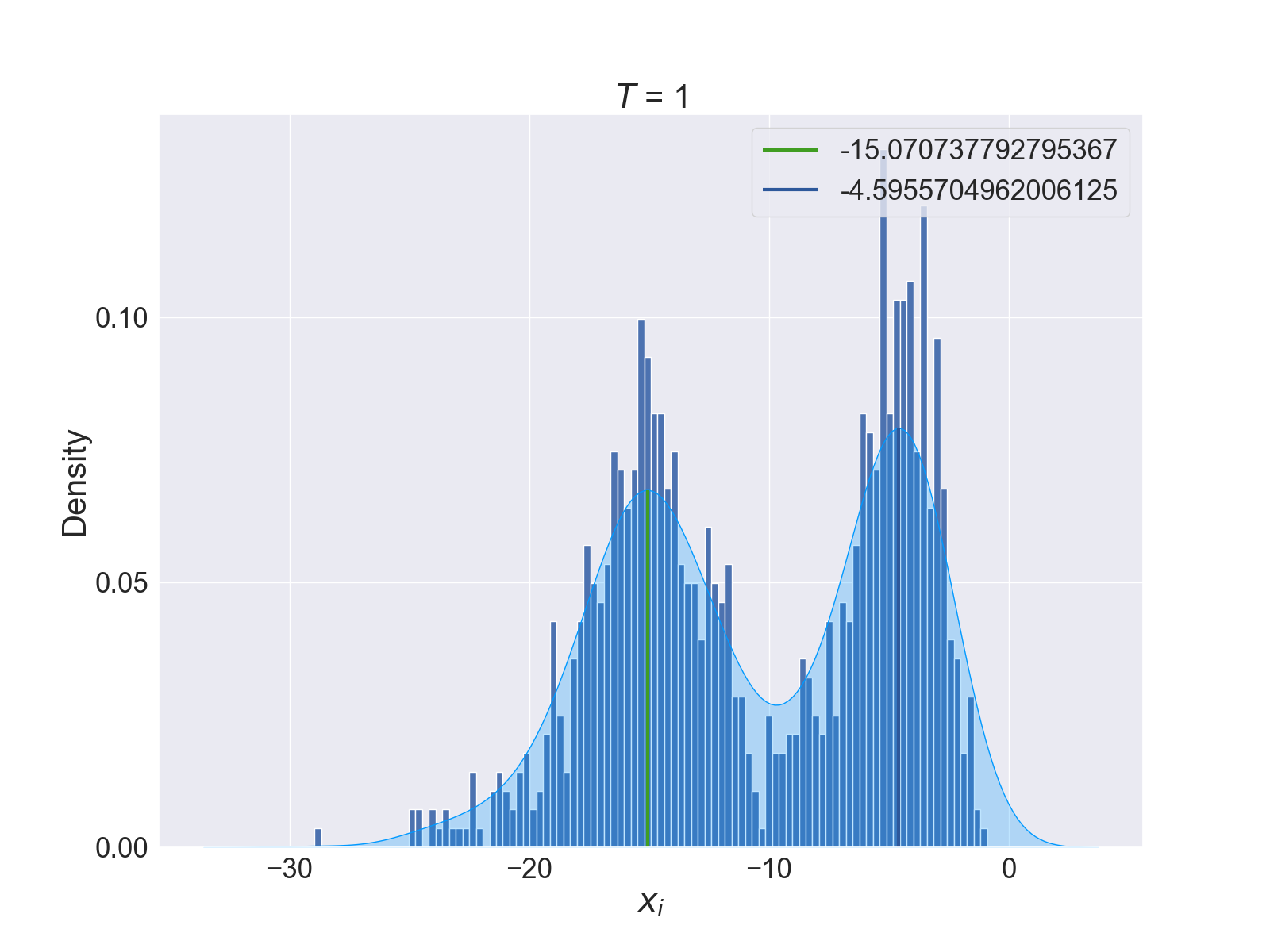}
         \caption{$T=1$}
         \label{fig:s1}
     \end{subfigure}
     \hfill
     \begin{subfigure}[b]{0.48\textwidth}
         \centering
         \includegraphics[width=\textwidth]{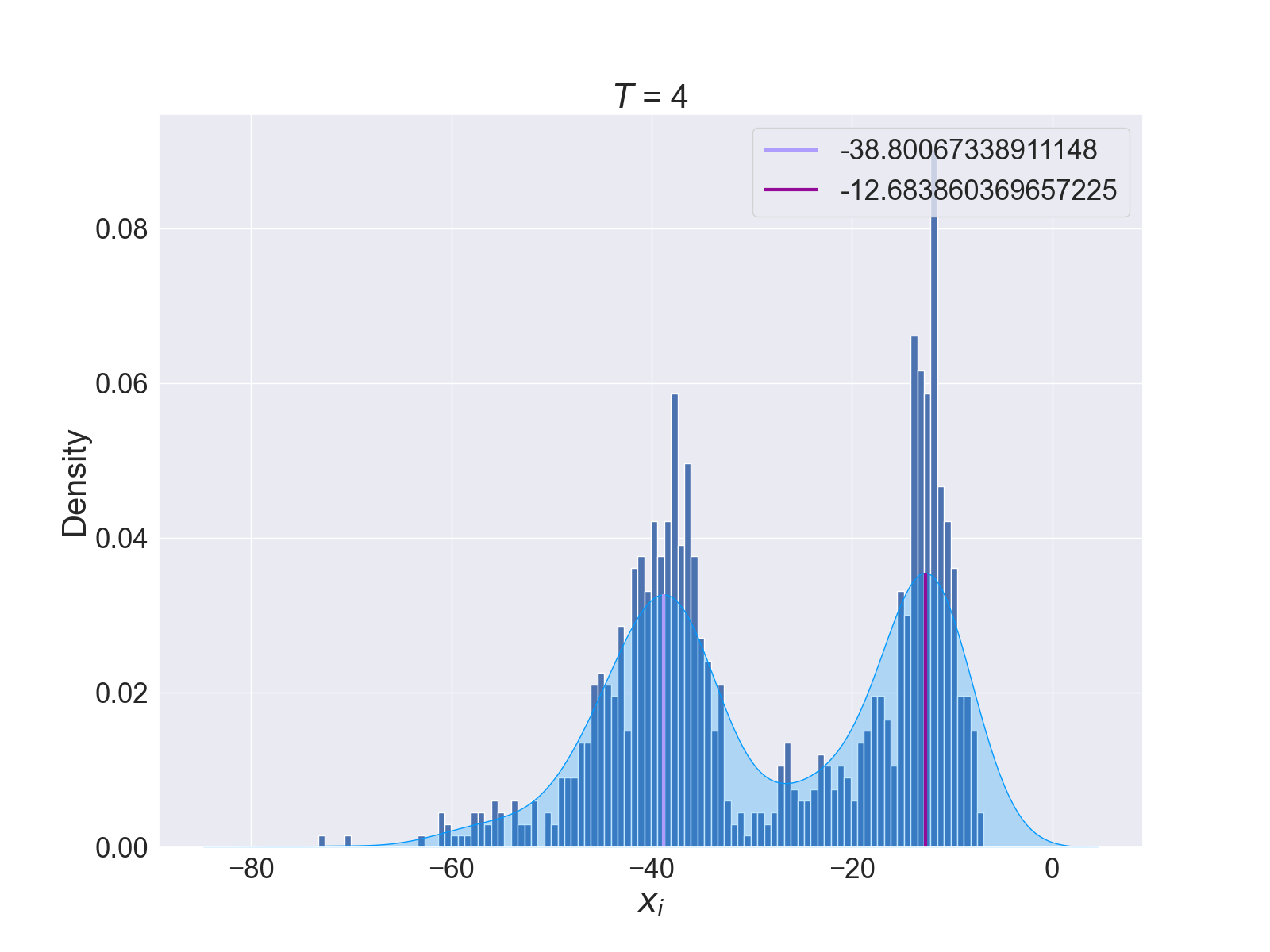}
         \caption{$T=4$}
         \label{fig:s2}
     \end{subfigure}
     \hfill
     \begin{subfigure}[b]{0.48\textwidth}
         \centering
         \includegraphics[width=\textwidth]{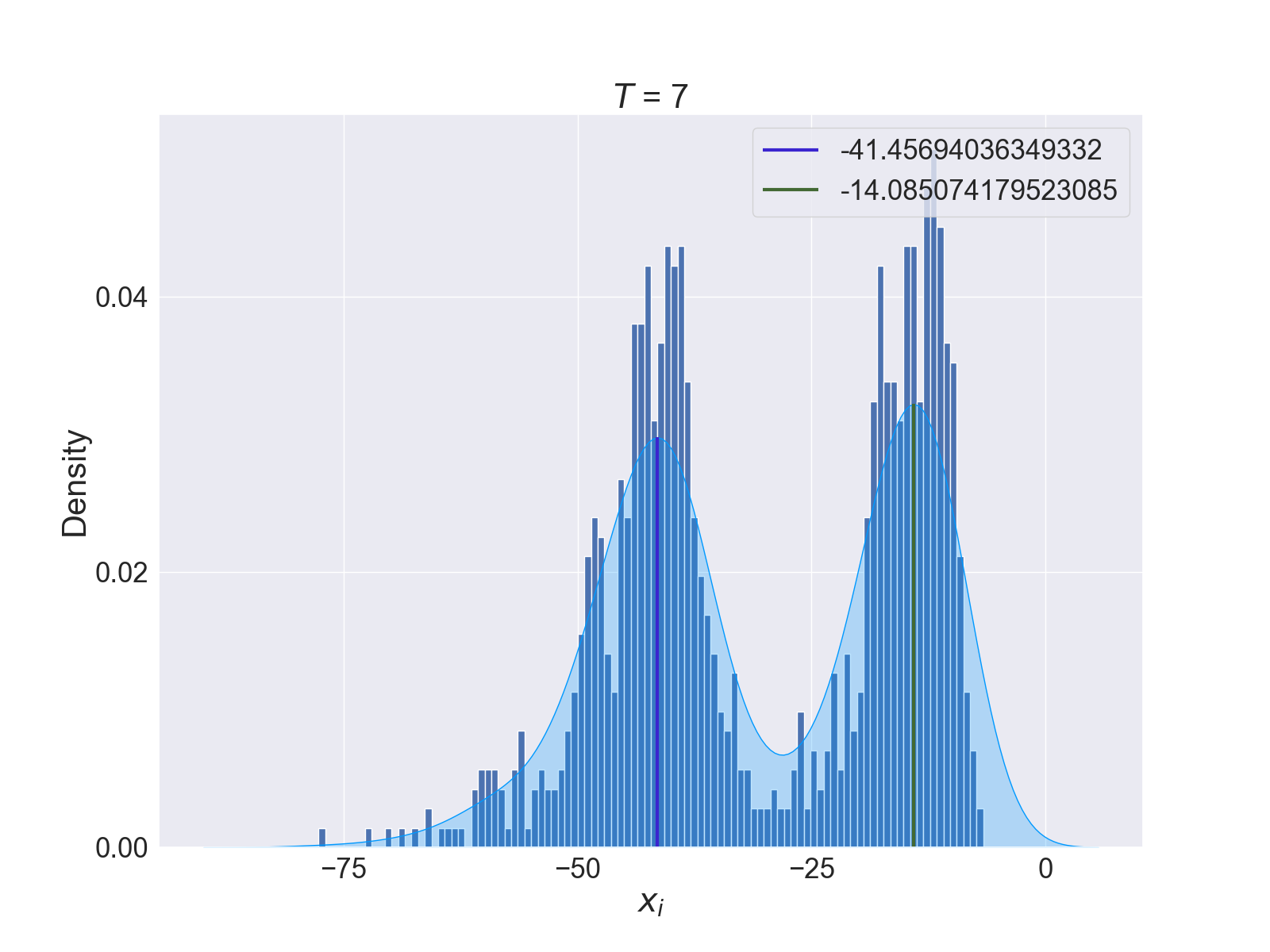}
         \caption{$T=7$}
         \label{fig:s3}
     \end{subfigure}
     \hfill
     \begin{subfigure}[b]{0.48\textwidth}
         \centering
         \includegraphics[width=\textwidth]{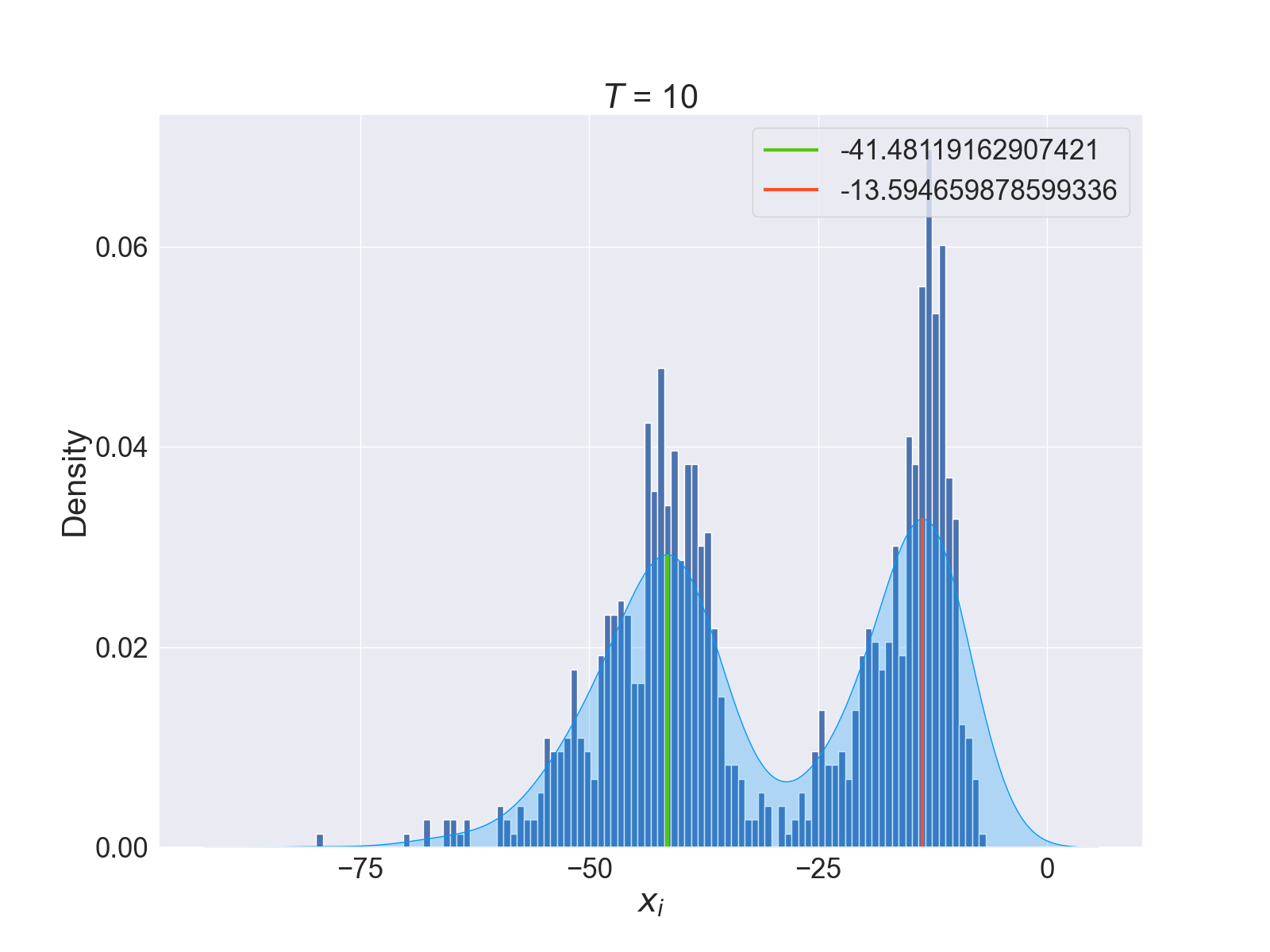}
         \caption{$T=10$}
         \label{fig:s4}
     \end{subfigure}
        \caption{Emergent opinion groups}
        \label{fig:6}
\end{figure}

In section \ref{sec:3.2}, we found the phenomenon called \textit{fragmentation}, which refers to the situations when agents are split into multiple groups. This differs from \textit{polarization}, because agents with the same stance can also be split up. We ran a simulation with $N=1000$, $\beta=1$, $\alpha=2$, $\rho=0.5$ and we discover fragmentation emerges. See Fig. \ref{fig:5}. Two obvious opinion groups appear. We show the emergence of the opinion group in Fig. \ref{fig:6}. We note the distribution of agents' opinions at time $T=1,4,7,10$ and mark the lead opinions (most common opinion at that time) using colored bars.
We find the formation of an opinion group arises early (at time $T=1$) and later the lead opinions become more and more radical over time. Once some opinions become viral locally, opinion groups start to form. But since we deploy agents on a static social network $S$, agents will directly interact with their neighbors and can only interact indirectly with agents who are not their nearest neighbors.
Thus, the opinion groups may be local and \textit{fragmentation} happens unless some globally viral opinions appear and a complete \textit{radicalization} may be reached. The transition between \textit{fragmentation} and \textit{radicalization} can be studied in the future.

Next let us focus on how individual agents change their stances. If an agent changes his stance at time $t+1$, assuming a Markovian process, we suppose the transition is just due to the impact at time $t$. We analyze the renegade mechanism from 3 aspects. If agent $i$ is positive (negative) at $t$ but turns negative (positive) at time $t+1$, we say agent $i$ is a renegade at time $t$. To figure out a transition at $t+1$, we analyze the situation at time $t$.
\begin{itemize}
    \item  Find renegades at every time and calculate the average opinion of their nearest neighbors $\left<x^{NN} \right>$.
    \item The most influential neighbor with a different stance.
    \item Difference of renegade's friend and foe numbers.
\end{itemize}

\begin{figure}
     \centering
     \begin{subfigure}[b]{0.48\textwidth}
         \centering
         \includegraphics[width=\textwidth]{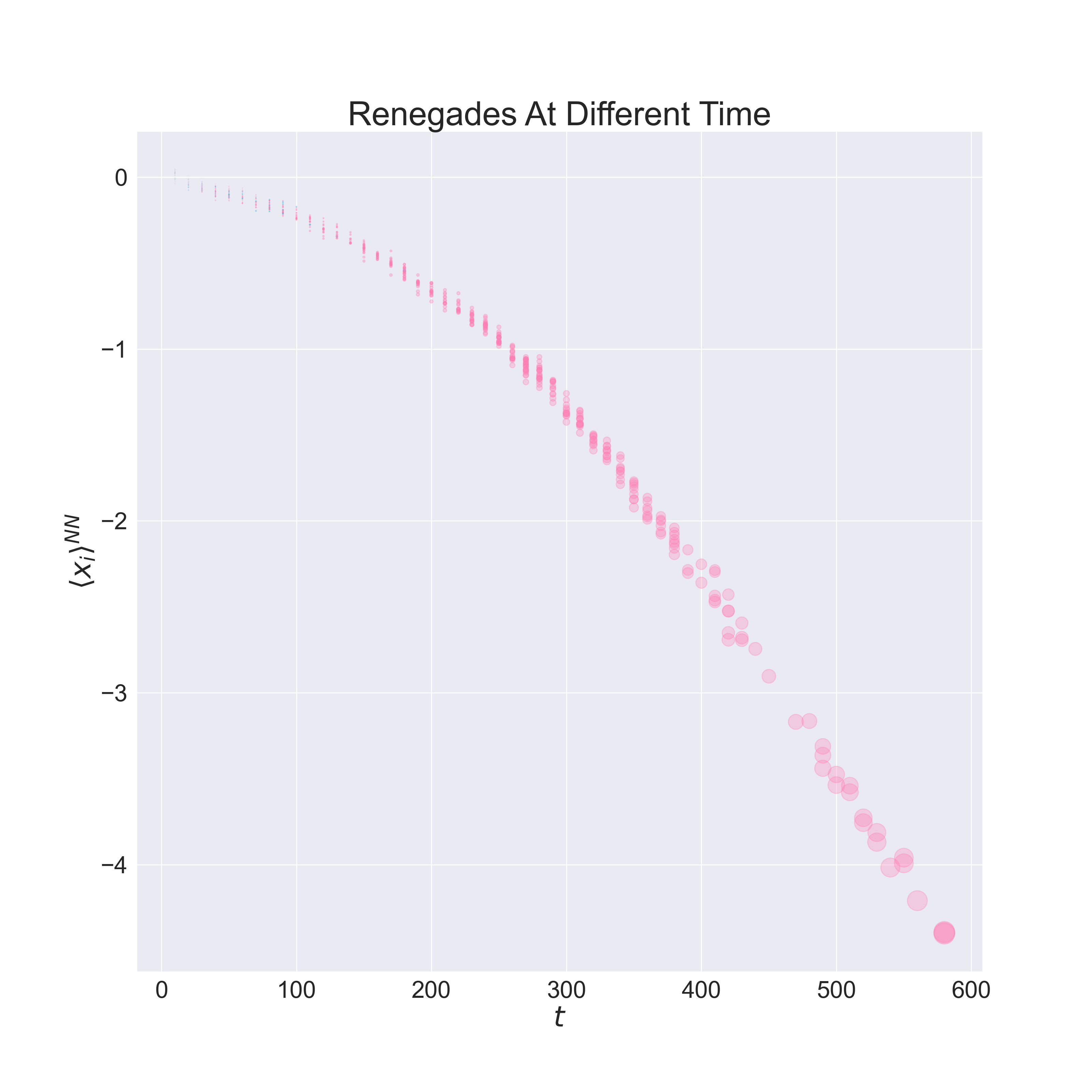}
         \caption{Renegades at Different Time}
         \label{fig:na}
     \end{subfigure}
     \hfill
     \begin{subfigure}[b]{0.48\textwidth}
         \centering
         \includegraphics[width=\textwidth]{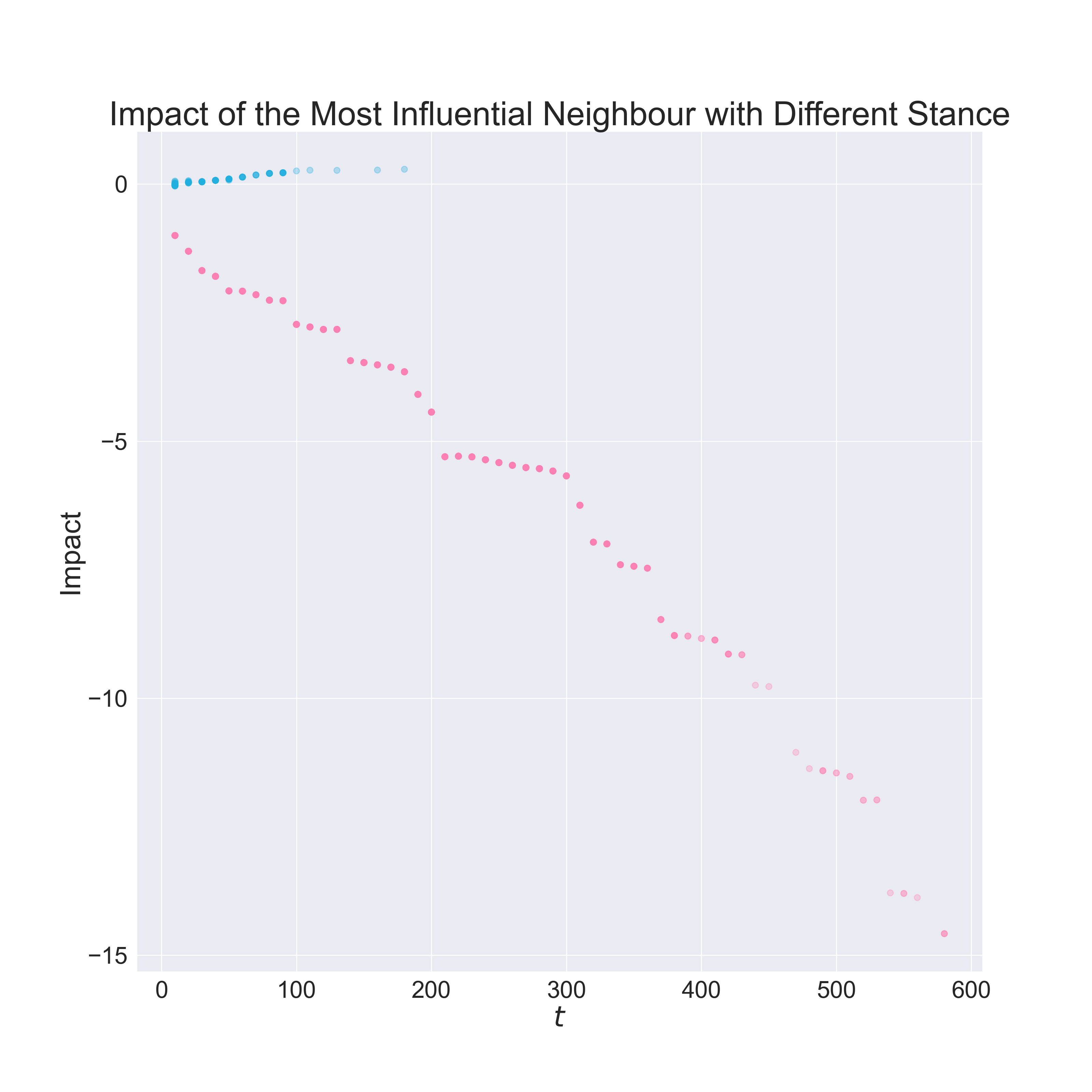}
         \caption{Most Influential Neighbour with Different Stance}
         \label{fig:nb}
     \end{subfigure}
     \hfill
     \begin{subfigure}[b]{0.48\textwidth}
         \centering
         \includegraphics[width=\textwidth]{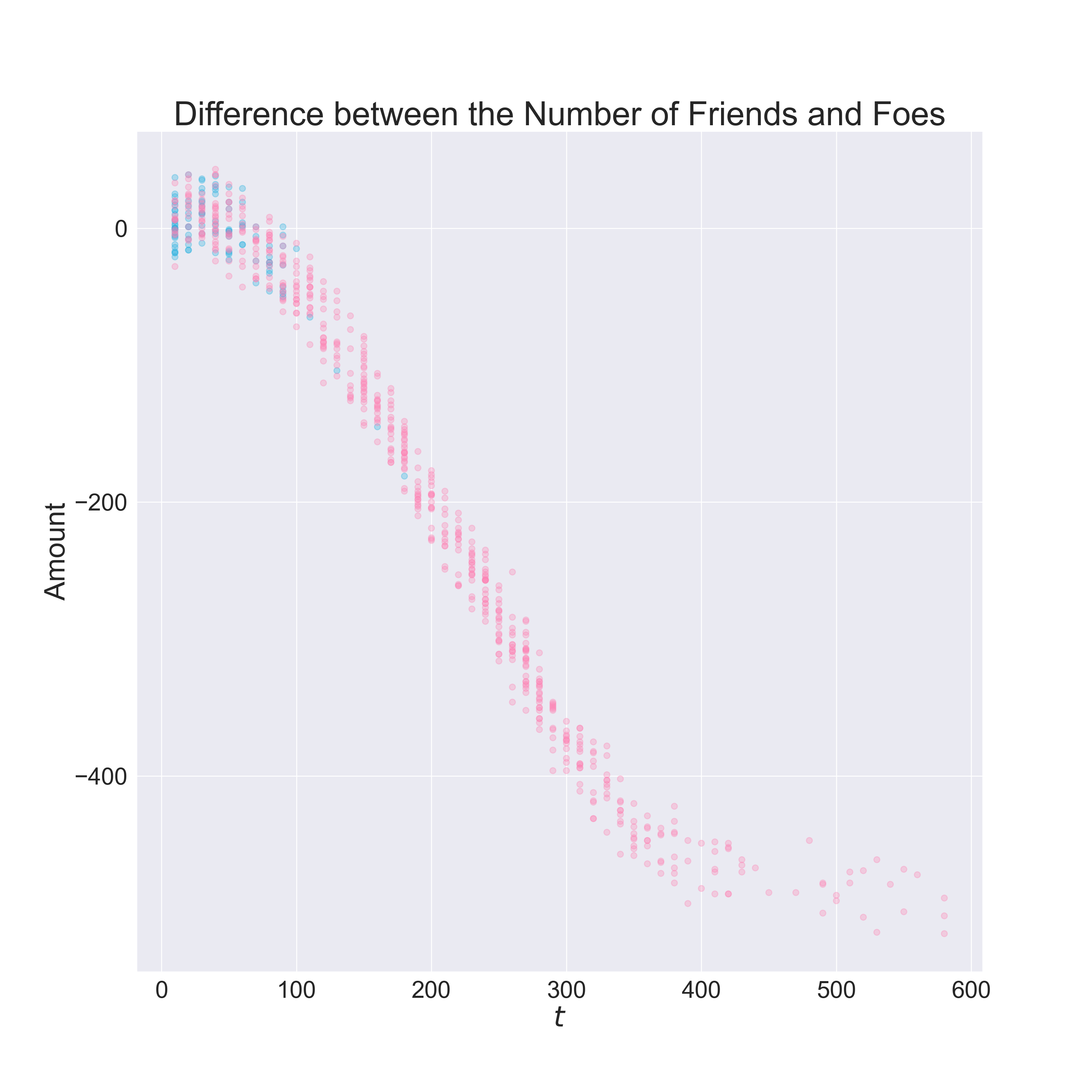}
         \caption{Difference between the Number of Friends and Foes}
         \label{fig:nc}
     \end{subfigure}
        \caption{The transition of Agent's Stance. We capture the results from the simulation Fig. \ref{fig:5} shows. If agent $i$ is positive (negative) at time $t$ and turns to negative (positive) at time $t+1$, we color him as pink (blue). Fig. \ref{fig:na}, renegades' nearest neighbor's average opinion $\left<x^{NN} \right>$ at different time steps. The area of dot denotes the relative value of $\left|\left<x^{NN} \right> \right|$.
        \ref{fig:nb}, the impact of the most influential neighbor with a different stance. Fig. \ref{fig:nc}, the number difference of friends and foes. To positive agents, if values at $y$ axis are larger than zero, friends are more than foes. Vice versa for the negative.}
        \label{fig:7}
\end{figure}

\begin{figure}
     \centering
     \begin{subfigure}[b]{0.48\textwidth}
         \centering
         \includegraphics[width=\textwidth]{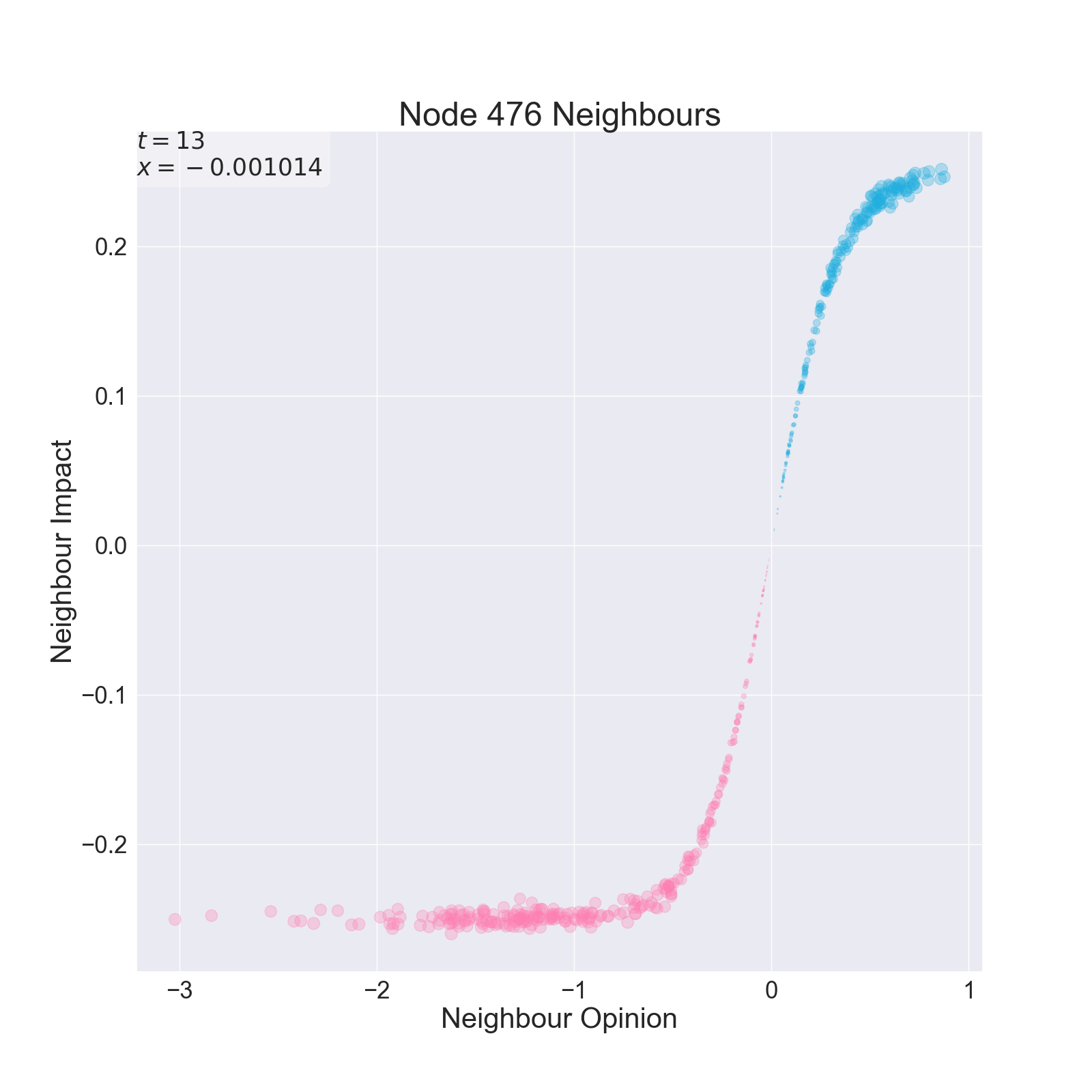}
         \caption{Agent 476}
         \label{fig:n7}
     \end{subfigure}
     \hfill
     \begin{subfigure}[b]{0.48\textwidth}
         \centering
         \includegraphics[width=\textwidth]{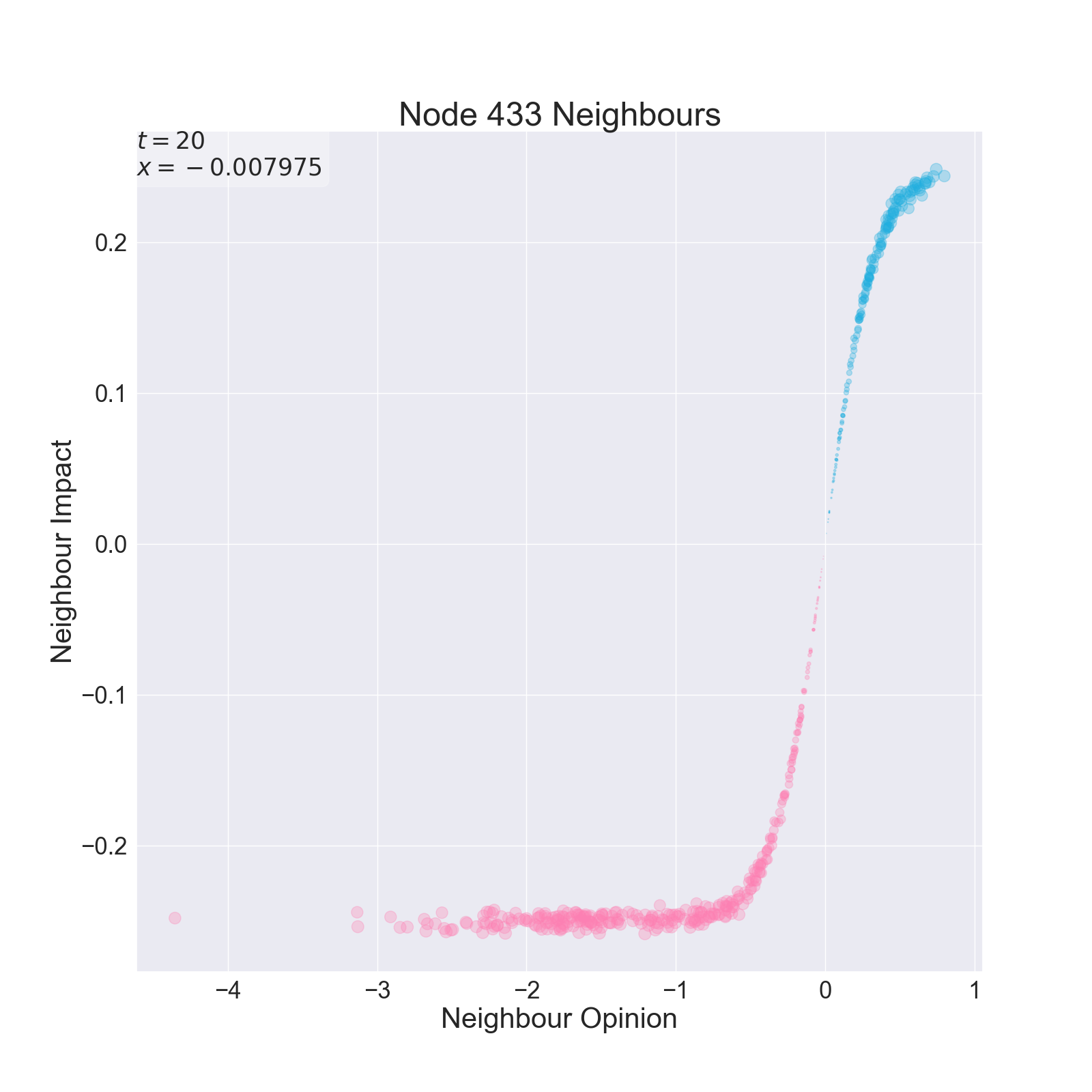}
         \caption{Agent 433}
         \label{fig:n8}
     \end{subfigure}
        \caption{Two specific anomalous agent's transitions}
        \label{fig:8}
\end{figure}

We reproduce the mechanism of simulation in Fig. \ref{fig:5}, shown in Fig.  \ref{fig:7}. As we can see, the crowd reaches a \textit{radicalization}. In Fig.\ref{fig:na}, most dots are pink, which means transitions from positive to negative are of dominant proportion. 
We are more interested in the anomaly, which can be counterintuitive. However, we also observe that anomalies mostly occur at the beginning of the simulation. 
At the beginning, $\left|x_i \right|$ is relatively small. Referring to Eq.  \ref{eq:(6)}, agents with opposed opinions have more opportunities to influence each other. This makes, say agent is negative at first and besieged by negative agents but turns positive later, possible. As time elapses, $\left|x_i \right|$ is growing larger, due to the effect of homophily (Eq. \ref{eq:(6)} suggests) , contradicting agents are harder to interact. Besides, the \textit{radicalization} to negativity has started, positive agents are surrounded by negative agents, and they may finally turn negative. Therefore we give an explanation to why our so-called anomalies mainly happen at the beginning but later, most transitions are from positivity to negativity. In Fig. \ref{fig:nb} and Fig. \ref{fig:nc}, we can also provide another explanation. Though some agents are encompassed by dissidents (like $x_i^{(t+1)}\times \left<x_{NN}^{(t)} \right><0$ and $\sum_{j=1}I(x_jx_i>0)-\sum_{j}I(x_jx_i<0)$) , some relatively more influential neighbors are around them to counteract and pull them back to their side. We show two specific anomalous agent's transition patterns in Fig. \ref{fig:8}, which shows the force of the negative and the positive are in a stalemate. But as $\left|x_{476} \right|$ and $\left|x_{433} \right|$ are both small when the transition happens, the anomaly still arises. Overall, this is in accordance with our explanation above.

\clearpage
\clearpage
\section{Conclusion}
This study innovates the opinion dynamics model proposed in \cite{baumann2020modeling} and introduces a new model. Unlike the original model, this study assumes that agents have stable social relationships that limit their interactions with people they have confirmed before.  Based on this assumption, this study develops a new model for networks with various structures. Thus, this study’s models have more flexibility and better proximity to reality than the original one that assumes a fully connected network in this context. As shown in Eq. \ref{eq:(7)}, this study considers network structures by incorporating node centrality in the formula. This study discards the parameter of social interaction strength because of its ambiguity.

This study examines opinion dynamics from both macro and micro perspectives, namely collective opinion patterns and individual opinion evolution. The new model can reproduce three main opinion patterns: consensus, radicalization, and polarization. Moreover, this study also observes the phenomenon of \textit{fragmentation}, which means agents can be divided into more than two groups and this kind of separation can occur among agents with the same stance. Then this study demonstrates how opinion patterns change according to parameters, issue controversialness, and network structure. Using a simple trick, this study reveals the transition from \textit{radicalization} to \textit{polarization}, which was overlooked by the original model. This study also calculates the analytical critical transition between \textit{consensus} and \textit{radicalization} under the mean-field assumption, which fits well with the simulation.

Regarding the evolution of individual opinions, we identify the limitation of the nonlinear function $tanh(x)$ and propose a new nonlinear function.  We investigate the phenomenon of \textit{fragmentation} by analyzing the opinion distribution of agents at every time step $T$ ($T=1,2,\cdots ,n$) and find that opinion groups form early and some leading opinions also change over time. Then we examine how an agent changes his or her stance from three aspects. We focus on anomalous conditions when agents are surrounded by dissidents but still change their opinions. We find that anomalies mainly occur at the beginning of temporal evolution. Finally, we randomly select two specific renegades and assess the influence exerted on them by their neighbors to verify our hypotheses.

Our study provides another paradigm for research of opinion dynamics. For the sake of simplicity, we only deployed our models on random ER networks. There may be still inconsistency when applied to other types of complex networks. We hope this study can shed light on future research on opinion dynamics, and serve as a fundamental for studying human behavior for scientists from a broad range of backgrounds.

\label{sec:Conclusion}

\section*{Author Contributions}
Yixiu Kong designed the research, Jiarui Dong performed the theoretical analysis, Yixiu Kong and Yi-Cheng Zhang designed the figures and Jiarui Dong performed the numerical experiments. All authors wrote the paper. 

\section*{Acknowledgements}
This project has received funding from Research Funds for the Central Universities from the Beijing University of Posts and Telecommunications under grant No. 505022019.

\bibliographystyle{unsrt}  
\bibliography{references.bib}






\end{document}